\documentclass[12pt]{iopart}
\usepackage{iopams,setstack}
\usepackage{amssymb,bm,graphicx,url,epsfig}
\usepackage[ansinew]{inputenc}
\usepackage[bookmarks=true]{hyperref}
\usepackage{bm}
\usepackage{color}
\usepackage[justification=raggedright,format=plain,indention=.3cm,font=small]{caption}
\usepackage{subcaption}
\usepackage{sidecap}

\usepackage[toc,page]{appendix}

\def\beq{\begin{equation}}
\def\eeq{\end{equation}}
\newcommand{\eqref}[1]{(\ref{#1})}

\begin{document}
\title[Instantons without equations of motion: semiclassics on a Riemann surface]{Instanton calculus without equations of motion: semiclassics from monodromies of a Riemann surface}

\author{Tobias Gulden$^1$, Michael Janas$^1$, and Alex Kamenev$^{1,2}$}

\address{$^1$Department of Physics, University of Minnesota, Minneapolis, MN 55455, USA}
\address{$^2$William I. Fine Theoretical Physics Institute, University of Minnesota, Minneapolis, MN 55455, USA}
\ead{kamenev@physics.umn.edu}

\begin{abstract}
Instanton calculations in semiclassical quantum mechanics rely on integration along trajectories which solve classical equations of motion. However in systems with higher dimensionality or complexified phase space these are rarely attainable. A prime example are spin-coherent states which are used e.g. to describe single molecule magnets (SMM). We use this example to develop instanton calculus which does not rely on explicit solutions of the classical equations of motion. Energy conservation restricts the complex phase space to a Riemann surface of complex dimension one, allowing to deform integration paths according to Cauchy's integral theorem. As a result,  the semiclassical actions can be evaluated without knowing  actual classical paths. Furthermore we show that in many cases such actions may be solely derived from monodromy properties of the corresponding Riemann surface and residue values at its singular points. 
As an example, we consider  quenching of tunneling processes in SMM by an applied magnetic field.
\end{abstract}

\maketitle

\section{Introduction}
\label{sec:introduction}
Semiclassical WKB method is a powerful tool to calculate tunneling amplitudes as well as matrix elements of higher order perturbation theory in quantum mechanics \cite{LandauLifshitz}. Energy levels and their tunneling splitting are determined by Bohr-Sommerfeld quantization and Gamow's formula which both rely on calculation of certain action integrals \cite{LandauLifshitz}. To this end one needs to solve the classical equations of motion which in one dimension is doable, but in higher dimensions or complexified phase space can be rather non-trivial, if at all attainable. Considering trajectories in complex phase space is necessary either for non-Hermitian Hamiltonians \cite{Bender2002,JoglekarBarnett} or in coherent states formalism with nonlinear integration measure \cite{GargKochetov,Kochetov,GargStone,Stone}.

In this paper we discuss the example of spin-coherent states which describe single molecule magnets (SMM). The total spin of the molecule $J$ takes the role of an inverse Planck constant, so for $J\gg1$ semiclassical treatment is applicable. The action in this system is $S=\int\frac{\bar{z}dz-zd\bar{z}}{1+\bar{z}z}$, where $z$ is the stereographic projection of the spin direction in $(\theta,\phi)$ and $\bar{z}$ its formal complex conjugate \cite{Stone}. In the case of two classically degenerate minima the quantum mechanical degeneracy is lifted by the presence of instantons. To find an instanton trajectory both  $\theta$ and $\phi$ have to be treated as complex variables, which is equivalent to treating the formally complex conjugated $(z,\bar{z})$ as two {\em independent} complex variables. This leads to a phase space of complex dimension two. Examples of such calculations in the context of SMM have been given in e.g. \cite{GargKochetov,Stone,KececiogluGarg,KececiogluGargB}. However the procedure is extensive and explicit solutions are known in a limited class of models.

Our goal is to develop a method to obtain the required action integrals without explicitly solving the classical equations of motion. Complex energy conservation restricts the phase space to one complex dimension which can be identified with a Riemann surface. Its genus and specific form is determined by the Hamiltonian. By Cauchy's integral theorem integration in one complex dimension does not depend on the actual path of integration, only its relative position with respect to the singularities and branch points is relevant. The contribution of the singularities can be calculated from their residue values. To calculate the branch point contribution we apply a concept called monodromy transformation. For a certain set of values of the moduli a branch cut collapses and the Riemann surface is degenerate. A monodromy transformation is an analytic transformation of the moduli around this singular set of values. In the end the Riemann surface returns to its initial state but the integration trajectories obtain additional contributions (those come from additional cycles around the handles of the Riemann surface). This fact strongly constrains the action integrals as analytic functions of the moduli (parameters of the Hamiltonian and energy itself). Here we show that in some examples there are sufficient constraints to fully determine the action integrals. This applies ideas that were developed in the context of the Seiberg-Witten solution of supersymmetric field theories \cite{SeibergWitten,SeibergWittenB,SWreview} and applied by the present authors \cite{TGMJPKAK,TGMJAK} in the context of statistical mechanics of 1D plasmas. In the following we will explain this concept in more detail.

The paper is organized as follows: section \ref{sec:SMM} presents basic facts about single molecule magnets and spin coherent states, and introduces the specific system that will be discussed below. In section \ref{sec:RiemannSurface} we show in detail how geometric reasoning on the Riemann surface allows to solve for the actions only by calculating residues and performing monodromy transformations. The results for energy states, level splitting and its characteristic oscillation with the applied magnetic field in SMM are compared with numerical calculations. Section \ref{sec:conclusion} gives a brief discussion of the results and the scope of the Riemann surface method.

\section{Single molecule magnets and spin-coherent states}
\label{sec:SMM}
Single-molecule magnets are large molecules with several metallic atoms. Their spins are fixed with respect to each other and thus at low temperatures act like one large spin \cite{LiudelBarcoHill,delBarco,Garanin,Gatteschi}. One of the most widely studied examples is $[Fe_8O_2(OH)_{12}(tacn)_6]^{8+}$, short name $Fe_8$, with a total spin $J=10$ \cite{ WernsdorferSessoli,Barra,Sangregorio,KececiogluGarg}. The anisotropy of the effective spin Hamiltonian can be derived directly from the molecule's symmetry properties \cite{LiudelBarcoHill}, in an external magnetic field up to leading order it is
\begin{equation}
 \hat{\mathcal{H}} = k_1J_x^2 + k_2J_y^2 - g\mu_B\vec{J}\cdot\vec{H}.
 \label{eq:H}
\end{equation}
In the usual notation $k_1>k_2>0$, therefore $x$ is hard, $y$ medium and $z$ the easy axis. Experimentally, $g\simeq2$, $k_1\simeq0.338K$ and $k_2\simeq0.246K$ \cite{WernsdorferSessoli}. It was shown \cite{KececiogluGarg} that for full quantitative understanding additional fourth-order anisotropy terms in \eqref{eq:H} need to be taken into account, but qualitative effects are similar.

\begin{figure}
 \includegraphics[width=\textwidth]{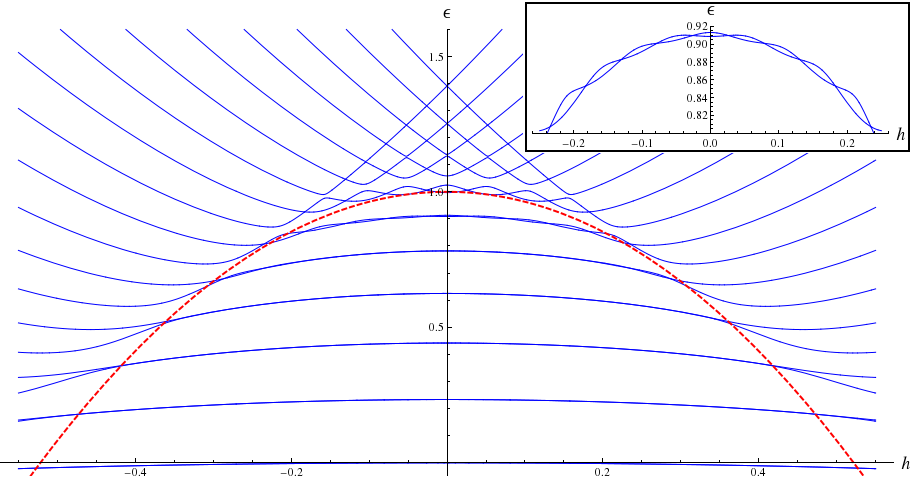}
 \caption{Energy levels for $\lambda=0.728$ and $J=10$ versus applied magnetic field. The red-dashed line marks the critical value for the field, underneath it pairs of levels are nearly degenerate. The splitting is caused by instantons and oscillates with applied field, the inset shows the example for the sixth pair of levels.}
 \label{fig:energies}
\end{figure}

Classically there are two degenerate ground states with the spin pointing along the $z$ axis in positive or negative direction. Quantum mechanically this degeneracy is lifted by magnetic quantum tunneling \cite{Sangregorio,WernsdorferSessoli}. However the splitting of the two levels oscillates with the strength of the external magnetic field $\vec{H}||\hat{x}$,\cite{WernsdorferSessoli} cf. figure \ref{fig:energies}. This is caused by the presence of two interfering instantons with complex actions. A magnetic field applied along the hard $x$ axis adds a Berry phase to the imaginary part, causing oscillations between constructive and destructive interference and thus of the level splitting \cite{Garg}. At zero field interference is purely constructive in a system with integer spin and purely destructive with half-integer spin. The latter implies that the two lowest states are perfectly degenerate, which is a manifestation of Kramer's theorem.

The $2J+1$ spin states are best described by SU(2) spin-coherent states $|z>=e^{zJ_+}|J,-J>$. Here $z$ is a stereographic coordinate of the spin direction, $\bar{z}$ its formal complex conjugate. However these have to be treated as two independent complex variables \cite{GargKochetov}. The instanton action in spin-coherent states is \cite{KececiogluGarg}
\begin{equation}
 S = -\int_\gamma \left(\frac{\dot{\bar{z}}z-\bar{z}\dot{z}}{1+\bar{z}z} - \mathcal{H}(z,\bar{z}) \right) dt,
 \label{eq:S}
\end{equation}
where
\begin{eqnarray}
 \hspace{-1.5cm}
 \mathcal{H}(z,\bar{z}) = \frac{<z|\hat{\mathcal{H}}|z>}{J<z|z>} = & k_1\left(J-\frac{1}{2}\right)\left(\frac{1-\bar{z}z}{1+\bar{z}z}\right)^2 + k_2\left(J-\frac{1}{2}\right)\frac{-(\bar{z}-z)^2}{(1+\bar{z}z)^2}\nonumber\\
 & + \frac{k_1+k_2}{2} - g\mu_BH\frac{-1+\bar{z}z}{1+\bar{z}z}
 \label{eq:energy-z}
\end{eqnarray}
is the expectation value of the Hamiltonian \eqref{eq:H}. The stereographic coordinates $(z,\bar{z})$ are taken along the $x$ axis, i.e. as projection onto a plane parallel to the $yz$-plane. The classical equations of motions are obtained by independent variations of this action with respect to $z$ and $\bar z$. The integration path $\gamma$ runs along a solution of these equations, satisfuing a proper boundary conditions.  Any such solution conserves the complex energy $\mathcal{H}$. As a result, the last term in the action (\ref{eq:S}) is always trivial and, in all cases considered below, is a pure  phase. The first term is the dynamic contribution. We treat it in the rest of the paper by developing a method to evaluate it without explicitly knowing the trajectory $\gamma$.

\section{Riemann surface calculations}
\label{sec:RiemannSurface}
In this section we  evaluate the action integrals on the Riemann surface and use the results to obtain energy levels and their tunneling splitting. To this end we define the new complex coordinates
\[
 p = \frac{1-z\bar{z}}{1+z\bar{z}};\quad \quad\quad q = \frac{z-\bar{z}}{i(z+\bar{z})}.
\]
The 1-form in equation \eqref{eq:S} is transformed into
\begin{equation}
 \sigma = -\frac{zd\bar{z}-\bar{z}dz}{1+\bar{z}z} = i(1-p)\frac{dq}{1+q^2},
 \label{eq:1-form}
\end{equation}
and complex energy conservation from equation \eqref{eq:energy-z} becomes
\begin{equation}
 \epsilon = \frac{1}{\lambda}(p-h)^2 + \frac{(1-p^2)q^2}{1+q^2},
 \label{eq:energy}
\end{equation}
where we defined $\lambda = k_2/k_1$, $h = \frac{-Hg\mu_B}{2k_1(J-1/2)}$ and $\epsilon = \frac{\mathcal{H} - (k_1+k_2)J/2}{k_2J(J-1/2)} + \frac{k_1h^2}{k_2}$. The shift in $\mathcal{H}$ fixes the classical minimum to $\epsilon=0$. Kinematics are restricted by energy conservation \eqref{eq:energy}, therefore all trajectories are confined to a Riemann surface $\mathcal{F}$ of complex dimension one inside the space of two complex dimensions $p,q$:
\begin{equation}
 \mathcal{F}(p,q) = ((p-h)^2-\epsilon\lambda)(1+q^2) + \lambda q^2(1-p^2) = 0.
 \label{eq:RS}
\end{equation}

\subsection{Analysis of the Riemann surface}
\label{sec:RSanalysis}
\begin{figure}
 \includegraphics[width=\textwidth]{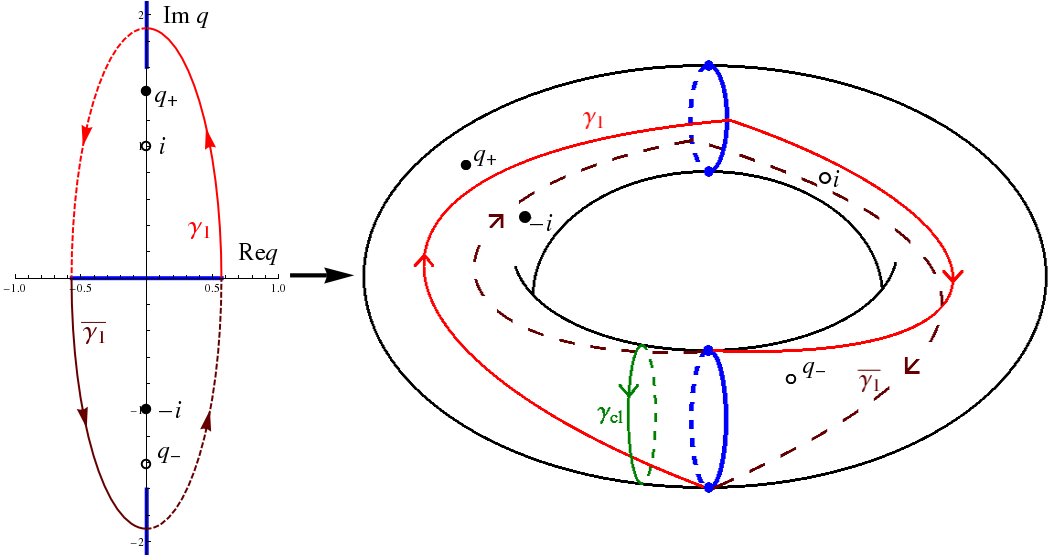}
 \caption{(Color online) Left: Plane of complex $q$ for $\lambda=0.5$, $h=0.4$ and $\epsilon=0.1$. The black points are the singularities $\pm i$ and $q_\pm$, where a full (open) point means the non-zero residue is only on the first (second) sheet. The blue lines are the branch cuts, the outer branch cut, along the imaginary axis, is continued to $\pm i\infty$. The instanton trajectories $\gamma_1$ ($\bar{\gamma}_1$) in light (dark) red connect two classical turning points (endpoints of inner branch cut) on opposite sheets, where a solid (dashed) line is on the first (second) sheet. Right: The Riemann surface of genus 1 can be seen as a torus. The green trajectory is the classical cycle around the inner branch cut. Combining $\gamma_1+\bar{\gamma}_1=\Gamma$ gives a closed instanton trajectory.}
 \label{fig:RS}
\end{figure}

To obtain the 1-form \eqref{eq:1-form} equation \eqref{eq:RS} is solved for $p=p(q)$:
\begin{equation}\hspace{-0.1\textwidth}
 p(q) = \frac{h(1+q^2) \pm \sqrt\lambda \sqrt{\epsilon+[\epsilon(2-\lambda)-1+h^2]q^2+[\epsilon-\epsilon\lambda+(h^2+\lambda-1)]q^4}}{(1+q^2-\lambda q^2)}.
 \label{eq:p}
\end{equation}
Substituting this into $\sigma$ in equation \eqref{eq:1-form} reduces the problem to integration in one complex variable $q$. The square root and $\pm$ sign in equation \eqref{eq:p} imply $p(q)$ is double valued, i.e. the Riemann surface \eqref{eq:RS} has two different sheets. Locally $p(q)$ is single valued and analytic except near the four zeroes of the square root, these are the branch points. To obtain a globally analytic function one connects the branch points pairwise via (arbitrarily chosen) branch cuts and performs analytic continuation of $p(q)$ by jumping to the other sheet of the Riemann surface whenever a cut is crossed. Figure \ref{fig:RS} shows the complex plane in $q$ with two branch cuts (blue). The cut around the origin along the real axis is labeled the inner branch cut, the cut along the imaginary axis which is closed through $\infty$ is labeled the outer branch cut. A Riemann surface with two branch cuts has genus $g=1$ and is topologically a torus.

Besides the branch points there are more special points on the Riemann surface, the singularities in $\sigma$. The measure $\frac{dq}{1+q^2}$ diverges at $q=\pm i$, $p(q)$ diverges at $q_\pm=\frac{\pm i}{\sqrt{1-\lambda}}$. The residues of $\sigma$ at these points are easily calculated, but the values differ on the two sheets. Without loss of generality the first sheet is identified so that at $q=0+i\delta$ the positive sign in \eqref{eq:p} is assumed and the square root is evaluated with positive real part, everywhere else the definition follows from analytic continuation. Under this definition the residues of $\sigma$ are
\begin{equation}\hspace{-0.1\textwidth}
 \mathrm{Res}_i^{(2)}(\sigma)=1, \quad \mathrm{Res}_{-i}^{(1)}(\sigma)=-1,
 \quad \mathrm{Res}_{q_+}^{(1)}(\sigma)=\frac{h}{\sqrt{1-\lambda}},  \quad \mathrm{Res}_{q_-}^{(2)}(\sigma)=\frac{-h}{\sqrt{1-\lambda}}.
 \label{eq:Residue}
\end{equation}
Here the superscript denotes the sheet on which the poles are on. On the respective other sheet the residues are zero, i.e. these are removable singularities. Therefore there is a total of four poles on the Riemann surface, not four poles per sheet. These are marked in figure \ref{fig:RS} in black.

\subsection{Evaluating action integrals}
\label{sec:calculating}
The trajectories live on a Riemann surface with complex dimension one, so by Cauchy's theorem any continuous deformation of the path of integration does not change the value of the integral. Practically this implies one does not need to know the precise path of integration, only its relative position to the branch points and poles.

There are two classically degenerate energy minima of the Hamiltonian \eqref{eq:H} which are mapped onto the same value $q=0$, but they are on opposite sheets of the Riemann surface. For non-zero energy spin precession around a classical minimum appears as oscillation around the origin, the inner branch points are the turning points. The classical trajectory goes along the inner branch cut. An instanton trajectory needs to connect two classical turning points on different sheets, i.e. it has to cross the outer branch cut. The two possibilities are $\gamma_1$, $\bar{\gamma}_1$ in figure \ref{fig:RS}.

\subsubsection{Energy levels}
The classical action $S_{cl}(\epsilon)=\int_{\gamma_{cl}}\sigma$ is analytic in $\epsilon$, thus it can be expanded in a Taylor series:
\begin{equation}
 S_{cl}(\epsilon) = S_{cl}(0) + S'_{cl}(0)\epsilon + \mathcal{O}(\epsilon^2) = \epsilon\oint_{\gamma_{cl}}\partial_\epsilon\sigma|_{\epsilon=0} + \mathcal{O}(\epsilon^2).
\end{equation}
and $\gamma_{cl}$ may be deformed into a closed trajectory around the inner branch cut. In the limit $\epsilon\to0$ both branch points coincide at $q=0$ and $\sigma$ is regular in the vicinity. Thus $S_{cl}(0)=0$ and all derivatives of $S_{cl}(\epsilon)$ only depend on residue values at $q=0$. For the first derivative
\begin{equation}
 \partial_\epsilon\sigma|_{\epsilon=0} = \frac{-i \sqrt\lambda dq}{2q\sqrt{1-h^2 + (1-h^2-\lambda)q^2}},
\end{equation}
and calculating the residue gives
\begin{equation}
 S_{cl}(\epsilon) = \frac{\pi\epsilon\sqrt\lambda}{\sqrt{1-h^2}}.
 \label{eq:S-cl}
\end{equation}
The quantization condition $S_{cl}=\frac{2\pi n}{J}$ for spin Hamiltonians in SU(2) \cite{Kochetov} determines the energy levels as 
\begin{equation}
 \epsilon_n = \frac{2n\sqrt{1-h^2}}{J\sqrt\lambda}.
 \label{eq:levels}
\end{equation}
Higher corrections can be easily calculated by evaluating residues of higher $\epsilon$ derivatives of $\sigma$ at $\epsilon=0$. Thus energy states can be evaluated by the use of a simple residue calculation.

\subsubsection{Oscillation period}
Two degenerate levels are split by tunneling which goes as
\begin{equation}
 \Delta\propto e^{-JS_1}+e^{-J\bar{S}_1} = e^{-J{\bf Re}S_1}Cos({\bf Im}S_1),
 \label{eq:Gamow}
\end{equation}
where $S_1$, $\bar{S}_1$ are the actions of the two instantons $\gamma_1,\bar{\gamma}_1$ in figure \ref{fig:RS} which are complex conjugated. The imaginary part of $S_1$ yields an oscillatory term, the real part an overall magnitude. For $\epsilon=0$ the only contribution to the imaginary part comes from the singularities on the Riemann surface. This can be seen when deforming $\gamma_1$ in figure \ref{fig:RS} to go from the origin along the imaginary axis. It passes two first-order poles at $q=i$ and $q=q_+$, which enter as half-residues when passing in one direction. Contour $\gamma_1$ passes them in counter-clockwise direction on their respective sheet, thus
\begin{equation}
 i{\bf Im}S_{1} = \pi i \left(\mathrm{Res}_i^{(2)}+\mathrm{Res}_{q_+}^{(1)}\right) = \pi i \left(1+\frac{h}{\sqrt{1-\lambda}}\right).
 \label{eq:Im-S1}
\end{equation}
From equation \eqref{eq:Gamow} it follows that instantons interfere destructively for $J{\bf Im} S_1=\frac{\pi}{2},\frac{3\pi}{2},...$, i.e. the condition for the tunneling splitting to be absent (so called "quenching", see Fig.~\ref{fig:energies})   is
\begin{equation}
 h = \sqrt{1-\lambda}\, \frac{m+\frac{1}{2}}{J},
 \label{eq:h-quench}
\end{equation}
for an integer $m$. This is in perfect agreement with numerical simulations (see figure \ref{fig:bandwidth}) and the behavior cited in literature \cite{Garg}. Here it is derived only out of geometric reasoning, without solving the equations of motion.

For excited levels $\epsilon\neq0$ there is an additional contribution to the imaginary part from integration along the real axis between the two turning points. The contribution to both $\gamma_1$ and $\bar{\gamma}_1$ is half the classical cycle $\gamma_{cl}$. Applying the quantization condition for the classical action shows that the additional contribution is ${\bf Im}S_{1} = \frac{\pi n}{J}$ for the $n$th level, and the same for $\bar{S}_1$. This causes a phase shift of $n\pi$ in the oscillatory part and leaves the level splitting \eqref{eq:Gamow} unchanged. The quenching condition \eqref{eq:h-quench} thus also holds for higher levels. Furthermore every trajectory with the same boundary conditions has the same ${\bf Im}S_1$ and a conjugated partner. Therefore condition \eqref{eq:h-quench} holds to all orders in the semiclassical expansion and is the full quantum mechanical quenching condition. Numerically this holds up to $10^{-15}$.

\subsubsection{Amplitude of the oscillations}
\label{sec:monodromy}
\begin{figure}[b]
 \includegraphics[width=\textwidth]{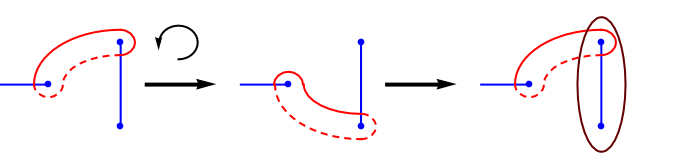}
 \caption{Schematics of a monodromy transformation where two branch points (blue) are exchanged. Left: original cycle (red), where the solid (dashed) part is on the first (second) sheet. Middle: The two branch points to the right exchanged clounter-clockwise. Right: Constructing the original cycle from the transformed cycle introduces an additional cycle around the branch cut (claret-red).}
 \label{fig:exchange}
\end{figure}

To obtain the real part of $S_1$ combine $\gamma_1+\bar{\gamma}_1$ in figure \ref{fig:RS} into a closed cycle $\Gamma$ with action $S_{in}=2{\bf Re}S_1$. Then the splitting \eqref{eq:Gamow} becomes $\Delta\propto e^{-JS_{in}/2}\cos({\bf Im}S_1)$. Straightforward integration along $\Gamma$ is rather tedious but in complex space an integral is uniquely defined by special points. On the Riemann surface $\mathcal{F}(p,q)$ in \eqref{eq:RS} two branch points collide to make $\mathcal{F}$ degenerate if for a point $(p,q)$ the two derivatives vanish $\partial_p\mathcal{F}=0$, $\partial_q\mathcal{F}=0$. This happens if the moduli of the Riemann surface $\mathcal{F}=0$, i.e. parameters $(\lambda,h,\epsilon)$, are 
\begin{equation}\hspace{-1cm}
 a)\; \epsilon=0, \quad b)\; \lambda=0, \quad c)\; (h\pm1)^2=\epsilon\lambda, \quad \mathrm{or}\quad d)\; 1-\lambda-h^2 = \epsilon(1-\lambda).
 \label{eq:critical}
\end{equation}
For all other values $S_{in}$ is analytic in the moduli. Note that condition $d)$ is the case for two branch points colliding at $q=\infty$. The physically relevant range is $0<\lambda<1$, small energy $\epsilon>0$ and fields $h$ below a critical value given by (\ref{eq:critical}d). Case (\ref{eq:critical}c) is beyond that value and therefore not relevant. 

\begin{figure}[t]
 \includegraphics[width=\textwidth]{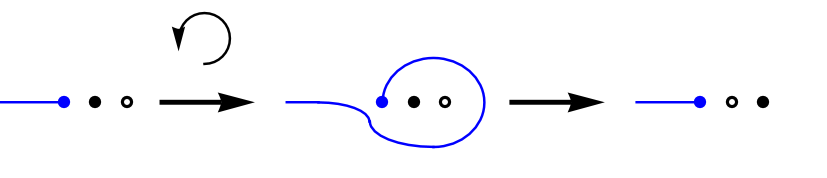}
 \caption{Schematics of a monodromy transformation where the branch cut (blue) entangles singularities (black). Left: original branch cut structure, where full (empty) circle is a singularity on first (second) sheet. Middle: The branch cut moved around the singularities clounter-clockwise. Right: The branch cut is brought to its original position. During this process both singularities cross the branch cut to the respective other sheet. To restore the initial singularities the notion of first and second sheet is inverted.}
 \label{fig:entangle}
\end{figure}

The non-analytic contributions can be identified by monodromy transformations, i.e. analytic continuation of the moduli around the critical values \eqref{eq:critical}. A full monodromy cycle returns the Riemann surface to its initial state, but branch points and singularities move and may exchange position. During a transformation the integration contour should  not cross a branch cut or singularity, it rather gets pulled along with the special points. The resulting cycle differs from the original trajectory  by an addition of an integer number of basic cycles around the handles of the Riemann surface. These additional cycles are the net change due to the monodromy transformation. Two main examples of monodromies are schematically shown in figures \ref{fig:exchange} and \ref{fig:entangle}.

First consider the transformation in only one of the moduli, $\lambda$. Therefore assume $\epsilon=0^+$ and $h=0^+$. Note that for $h=0$ the outer branch points coincide with the singularities $q_\pm$, for $\epsilon=0$ the inner branch cut collapses into a point at $q=0$ (cf. figure \ref{fig:RS}).

\paragraph{$\lambda=0$, (\ref{eq:critical}b)} Rotating $\lambda\to\lambda e^{2\pi i}$ rotates the outer branch points around the neighboring pair of singularities in counter-clockwise direction. The cycle $\Gamma$ is not affected by this. However the outer branch cut winds around the singularities as schematically shown in figure \ref{fig:entangle}. Restoring the original branch cut and singularities inverts the notion of first and second sheet, implying that now $\Gamma$ is on the opposite sheet. Compared to $\Gamma$ being on its initial sheet this introduces an overall minus sign:
\begin{equation}
 S_{in} \to -S_{in} \quad \mathrm{for} \quad \lambda\to\lambda e^{2\pi i}
 \label{eq:lambda0}
\end{equation}
Therefore $S_{in}$ is an odd function of $\sqrt{\lambda}$. This implies that $\sqrt\lambda$ is rather to be seen as the modulus of the Riemann surface than $\lambda$.

\paragraph{$\lambda=1$, (\ref{eq:critical}d)} In terms of $\sqrt\lambda$ there are two possible monodromies, $(1\pm\sqrt\lambda)\to(1\pm\sqrt\lambda)e^{2\pi i}$. In both cases the two outer branch points exchange position in clockwise direction and $\Gamma$ obtains two additional cycles similar to figure \ref{fig:exchange}. Both go around the inner branch points and all singularities, one clockwise on the first sheet, one counter-clockwise on the second sheet. Only the singularities $\pm i$ contribute with their residue values \eqref{eq:Residue} to yield 
\begin{eqnarray}
 S_{in} \to S_{in} + 2\cdot2\pi i \quad \mathrm{for} \quad (1-\sqrt\lambda)\to(1-\sqrt\lambda)e^{2\pi i},\nonumber\\
 S_{in} \to S_{in} - 2\cdot2\pi i \quad \mathrm{for} \quad (1+\sqrt\lambda)\to(1+\sqrt\lambda)e^{2\pi i}.
 \label{eq:lambda1trans}
\end{eqnarray}
Every repeated monodromy cycle causes the same change. The only function that constantly adds the same amount when the phase of the argument is changed is the complex logarithm. Therefore $S_{in}(\sqrt\lambda)$ needs to contain logarithmic terms as
\begin{eqnarray}
 S_{in} = Q_+ + 2\ln(1-\sqrt\lambda),\nonumber\\
 S_{in} = Q_- - 2\ln(1+\sqrt\lambda),
 \label{eq:lambda1}
\end{eqnarray}
where $Q_+$ ($Q_-$) is analytic near $\sqrt\lambda=1$ ($\sqrt\lambda=-1$).

From these conditions one can obtain $S_{in}$ by identifying $Q_\pm$. To this end define $g_\pm(\sqrt\lambda)$ via
\begin{equation}\hspace{-0.1\textwidth}
 Q_+(\sqrt\lambda) = -2\ln(1+\sqrt\lambda) + g_+(\sqrt\lambda), \quad\quad Q_-(\sqrt\lambda) = +2\ln(1-\sqrt\lambda) + g_-(\sqrt\lambda).
 \label{eq:Q}
\end{equation}
$g_\pm$ are entire functions in $\sqrt\lambda$. Equating the two expressions for $S_{in}$ in \eqref{eq:lambda1} shows $g_+(\sqrt\lambda) = g_-(\sqrt\lambda) = g(\sqrt\lambda)$. Similarly equation \eqref{eq:lambda0} can be used to show $g(\sqrt\lambda) = -g(-\sqrt\lambda)$. Thus $g(0)=0$ with a well-defined limit. Furthermore in equation \eqref{eq:energy} a change of $\lambda\to1/\lambda$ only is an exchange of coordinate axis $x\leftrightarrow y$. Therefore $g(\sqrt\lambda)$ is analytic and finite at $\infty$, i.e. $g$ is entire and bounded everywhere so by Liouville's theorem $g(\sqrt\lambda)=const=g(0)=0$. This yields for the instanton action
\begin{equation}
 S_{in} = 2\ln\left(\frac{1-\sqrt{\lambda}}{1+\sqrt{\lambda}}\right) = 2{\bf Re}S_1.
 \label{eq:Sin}
\end{equation}
This gives  the ground state splitting in the zero field limit \cite{Garg}.

Next consider the case with an applied magnetic field $h\neq0$. This separates the outer branch points from the singularities $q_\pm$ which obtain non-zero residue \eqref{eq:Residue}.
\newline {$h\to-h$}, Inverting the sign of $h$ is equivalent to changing the direction of the magnetic field, however the symmetry in \eqref{eq:H} requires that this leaves the physics unchanged. Therefore
\begin{equation}
 S_{in}(-h) = S_{in}(h).
 \label{eq:h0}
\end{equation}

\paragraph{$\lambda=0$, (\ref{eq:critical}b)} The monodromy effect is the same as for $h=0$, therefore equation \eqref{eq:lambda0} still holds and $S_{in}$ is an odd function in $\sqrt\lambda$.

\begin{figure}
 \includegraphics[width=\textwidth]{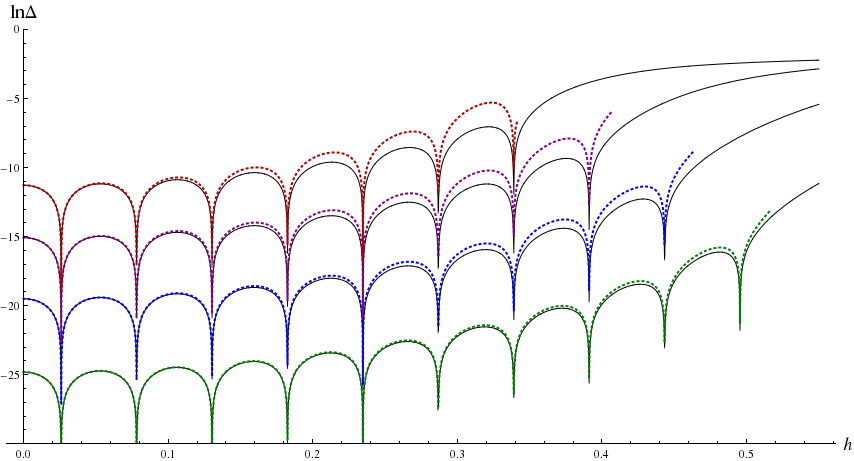}
 \caption{Analytic results (dashed, different colors) for the logarithm of the level splitting, $\ln\Delta$, compared to numerical calculations (solid-black). The preexponential factor in \eqref{eq:Gamow} was assumed to be constant and adjusted to have perfect agreement at $h=0$. The oscillation period of the quenchings is equal for all levels and agrees perfectly with numerical results. The amplitude agrees well, discrepancies mostly come from the approximated value for the energy where an overestimation of $\Delta$ agrees with the overestimation of $\epsilon$ in figure \ref{fig:energies} (the second-order expansion of $S_{cl}$ in $\epsilon$ was used), and $h$-dependence of the preexponential factor. Here $\lambda=0.728$ for the $Fe_8$ molecule \cite{WernsdorferSessoli}.}
 \label{fig:bandwidth}
\end{figure}

\paragraph{$1-\lambda-h^2=0$, (\ref{eq:critical}d)} There are two possibilities to perform this monodromy in $\sqrt\lambda$, namely to rotate around $\pm\sqrt{1-h^2}$. The effect is the same as for $h=0$, $\Gamma$ picks up cycles around the inner branch cut and all singularities. The net difference to \eqref{eq:lambda1} comes from the singularities $q_\pm$ which obtain non-zero residues:
\begin{eqnarray}
 S_{in} = Q_+ + 2 \left(1-\frac{h}{\sqrt{1-\lambda}}\right) \ln(\sqrt{1-h^2}-\sqrt\lambda),\nonumber\\
 S_{in} = Q_- - 2 \left(1-\frac{h}{\sqrt{1-\lambda}}\right) \ln(\sqrt{1-h^2}+\sqrt\lambda).
 \label{eq:lambda1h}
\end{eqnarray}
Alternatively (\ref{eq:critical}d) can be written as $1-\lambda-h^2 = (1-h^2)(1-\lambda)-h^2\lambda = 0$, and the monodromy is performed around $\sqrt{(1-h^2)(1-\lambda)}\pm h\sqrt\lambda = 0$. This notion gives the same transformation for the branch points and cycles, the equivalent condition for the action becomes
\begin{eqnarray}
 S_{in} = \tilde{Q}_+ + 2 \left(1-\frac{h}{\sqrt{1-\lambda}}\right) \ln(\sqrt{1-h^2}\sqrt{1-\lambda}-h\sqrt\lambda),\nonumber\\
 S_{in} = \tilde{Q}_- - 2 \left(1-\frac{h}{\sqrt{1-\lambda}}\right) \ln(\sqrt{1-h^2}\sqrt{1-\lambda}+h\sqrt\lambda).
 \label{eq:lambda1halt}
\end{eqnarray}
Again $S_{in}$ is identified by a similar procedure as in equations \eqref{eq:Q} and \eqref{eq:Sin}. It is easy to see that the term with the constant residue $1$ obeys the symmetry properties \eqref{eq:lambda0} and \eqref{eq:h0} as written in \eqref{eq:lambda1h}, the term with residue $h/\sqrt{1-\lambda}$ obeys these as written in equation \eqref{eq:lambda1halt}. Putting these observations together one finds
\begin{equation}\hspace{-0.05\textwidth}
 S_{in} = 2 \ln\left(\frac{\sqrt{1-h^2}-\sqrt{\lambda}}{\sqrt{1-h^2}+\sqrt{\lambda}}\right) - \frac{2h}{\sqrt{1-\lambda}}\ln\left(\frac{\sqrt{1-h^2}\sqrt{1-\lambda}-h\sqrt{\lambda}}{\sqrt{1-h^2}\sqrt{1-\lambda}+h\sqrt{\lambda}}\right). 
 %= 2{\bf Re}S_1.
 \label{eq:Sin-h}
\end{equation}
In the limit $h\to0$ \eqref{eq:Sin-h} reduces to \eqref{eq:Sin}, as required, and the resulting level splitting is in a perfect agreement with literature \cite{Garg} and numerical results for the lowest level in figure \ref{fig:bandwidth} (green).

With the same procedure $S_{in}$ can be found for excited states $\epsilon\neq0$. The relevant critical moduli are (\ref{eq:critical}a), (b) and (d), while (\ref{eq:critical}c) is outside the range of parameters.

\paragraph{$\epsilon=0$, (\ref{eq:critical}a)} Rotating $\epsilon$ around zero exchanges the two inner branch points in counter-clockwise direction (see example in \ref{fig:exchange}). As net effect, two cycles around the inner branch cut are added to $\Gamma$, clockwise on the first sheet and counter-clockwise on the second sheet, i.e. $\Gamma\to\Gamma-2\gamma_{cl}$. To obtain an additional term of twice the quantized classical action for every monodromy, $S_{in}$ contains a logarithmic term as
\begin{equation}
 S_{in} = Q_\epsilon - \frac{\epsilon\sqrt\lambda}{\sqrt{1-h^2}}\ln\epsilon,
 \label{eq:epsilon0}
\end{equation}
where expression \eqref{eq:S-cl} for the classical action was used.

\paragraph{$\lambda=0$, (\ref{eq:critical}b)} Additionally to the previously mentioned change of the sign $\Gamma$ gets twisted around the inner branch cut and obtains additional contributions of classical cycles $\Gamma\to-\Gamma-2\gamma_{cl}$, thus
\begin{equation}
 S_{in} \to -S_{in} - 2S_{cl} \quad \mathrm{for} \quad \lambda\to\lambda e^{2\pi i}.
\end{equation}
Therefore
\begin{equation}
 S_{in} = Q_\lambda - \frac{\epsilon\sqrt\lambda}{\sqrt{1-h^2}}\ln\lambda,
 \label{eq:lambda0-epsilon}
\end{equation}
and $Q_\lambda$ is antisymmetric in $\sqrt\lambda$.

\paragraph{$1-\lambda-h^2-\epsilon+\epsilon\lambda=0$, (\ref{eq:critical}d)} The net effect of this monodromy is the same as mentioned above equation \eqref{eq:lambda1trans}. For this monodromy $\sqrt\lambda$ rotates around $\pm\sqrt{1-h^2-\epsilon+\epsilon\lambda}$, the net change to equation \eqref{eq:lambda1h} is the contribution from the classical cycle:
\begin{eqnarray}\hspace{-0.05\textwidth}
 S_{in} = Q_+ + 2 \left(1-\frac{h}{\sqrt{1-\lambda}}-\frac{\epsilon\sqrt\lambda}{\sqrt{1-h^2}}\right) \ln(\sqrt{1-h^2-\epsilon+\epsilon\lambda}-\sqrt\lambda),\nonumber\\\hspace{-0.05\textwidth}
 S_{in} = Q_- - 2 \left(1-\frac{h}{\sqrt{1-\lambda}}-\frac{\epsilon\sqrt\lambda}{\sqrt{1-h^2}}\right) \ln(\sqrt{1-h^2-\epsilon+\epsilon\lambda}+\sqrt\lambda).
 \label{eq:lambda1epsilon}
\end{eqnarray}
Or alternatively
\begin{eqnarray}\hspace{-0.05\textwidth}
 S_{in} = \tilde{Q}_+ + 2 \left(1-\frac{h}{\sqrt{1-\lambda}}-\frac{\epsilon\sqrt\lambda}{\sqrt{1-h^2}}\right) \ln(\sqrt{1-h^2-\epsilon}\sqrt{1-\lambda}-h\sqrt\lambda),\nonumber\\\hspace{-0.05\textwidth}
 S_{in} = \tilde{Q}_- - 2 \left(1-\frac{h}{\sqrt{1-\lambda}}-\frac{\epsilon\sqrt\lambda}{\sqrt{1-h^2}}\right) \ln(\sqrt{1-h^2-\epsilon}\sqrt{1-\lambda}+h\sqrt\lambda).
 \label{eq:lambda1epsilonalt}
\end{eqnarray}
Now we use the same arguments as below Eq.~(\ref{eq:lambda1halt}) to identify the instanton action of the $n$th level as
\begin{eqnarray}\hspace{-0.15\textwidth}
 S_{in} = &-& \frac{\epsilon_n\sqrt\lambda}{\sqrt{1-h^2}}\ln\left(\epsilon_n\lambda(1-h^2-\epsilon_n+\epsilon_n\lambda-\lambda)\right) 
           + 2 \ln\left(\frac{\sqrt{1-h^2-\epsilon_n+\epsilon_n\lambda}-\sqrt{\lambda}}{\sqrt{1-h^2-\epsilon_n+\epsilon_n\lambda}+\sqrt{\lambda}}\right)\nonumber\\
          \hspace{-0.1\textwidth}&-& \frac{2h}{\sqrt{1-\lambda}}\ln\left(\frac{\sqrt{1-h^2-\epsilon_n}\sqrt{1-\lambda}-h\sqrt{\lambda}}{\sqrt{1-h^2-\epsilon_n}\sqrt{1-\lambda}+h\sqrt{\lambda}}\right).
        %   = 2{\bf Re}S_1.
 \label{eq:Sin-full}
\end{eqnarray}
Here $\epsilon_n$ is the energy of the $n$th level, cf. equation \eqref{eq:levels}. This is the full result for the instanton action of excited levels, which agrees with the previous result \eqref{eq:Sin-h} for the ground state $\epsilon_0=0$. Figure \ref{fig:bandwidth} compares the resulting splitting of the first three excited levels (blue, purple, red) to numerical results. For each curve the preexponential factor in \eqref{eq:Gamow} was chosen constant such that for $h=0$ the analytic and numerical results agree. The analytic result follows the simulated values well. The prefactor of the first logarithm comes from first-order approximation to $S_{cl}$ in equation \eqref{eq:S-cl}. Applying the quantization condition in SU(2) allows to replace $\frac{\epsilon_n\sqrt\lambda}{\sqrt{1-h^2}}=\frac{2n}{J}$ and make the result precise to higher orders.

\section{Discussion and conclusion}
\label{sec:conclusion}
Semiclassical calculations for large-spin systems in SU(2) involve instantons in complex phase space. Here we demonstrated that the necessary action integrals can be evaluated without explicitely knowing the trajectories that solve the equations of motion. It is sufficient to know which homotopy class in the fundamental group of the given Riemann surface these belong to, which can be obtained directly from geometric reasoning. Action integrals are evaluated purely from the behavior of special points like singularities and branch points. One only needs to consider residue values and monodromy transformations near the critical values of the moduli where the Riemann surface is degenerate. This allows to obtain energy levels for the given Hamiltonian \eqref{eq:H} and extend previous results for the level splitting to excited states with non-zero energy and an explicit proof that the oscillation period with the applied magnetic field does not change for higher levels and indeed extends to the full quantum mechanical result. The semiclassical results fit well with numerical simulations.

The method described in this paper is not limited to large-spin systems. In general it may be applicable to any semiclassical system where instanton trajectories are not fixed to real phase space. One example are non-Hermitian Hamiltonians which were studied intensively \cite{Bender2002,Bender2003,JoglekarBarnett,JoglekarSaxena} and their trajectories were shown to have rich and elaborate behavior \cite{ Nanayakkara,BenderClassTraj}. In \cite{TGMJPKAK,TGMJAK} the present authors show application to a class of non-Hermitian Hamiltonians that was derived from statistical mechanics. A second case are curved spaces with non-trivial measure of integration, like for spins in SU(2) or SU(1,1). In these cases geometric reasoning allows to perform instanton calculations without direct integration or without solving the classical equations of motion.

\section{Acknowledgement}
We want to thank Peter Koroteev for many useful discussions. 
%and his enthusiasm that accompanied us throughout the project. 
This work was supported by NSF grant DMR1306734.


\begin{thebibliography}{10}

\bibitem{LandauLifshitz}
L. Landau and E. Lifshitz.
\newblock {\em Quantum Mechanics: non-relativistic theory}, Pergamon Press, New York, 1977.

\bibitem{Bender2002}
C. M. Bender, D. C. Brody and H. F. Jones.
\newblock {\em Phys. Rev. Lett.}, 89:270401, 2002.

\bibitem{JoglekarBarnett}
Y. N. Joglekar and J. L. Barnett.
\newblock {\em Phys. Rev. A}, 84:024103, 2011.

\bibitem{GargKochetov}
A. Garg, E. Kochetov, K.-S. Park and M. Stone.
\newblock {\em Jour. of Math. Phys.}, 44:1, 2003.

\bibitem{Kochetov}
E. Kochetov.
\newblock {\em Jour. of Math. Phys.}, 36:366, 1995.

\bibitem{GargStone}
A. Garg and M. Stone.
\newblock {\em Phys. Rev. Lett.}, 92:1, 2004.

\bibitem{Stone}
M. Stone and K.-S. Park and A. Garg.
\newblock {\em Jour. of Math. Phys.}, 41:12, 2000.

\bibitem{KececiogluGarg}
E. Ke\ifmmode \mbox{\c{c}}\else \c{c}\fi{}ecio\ifmmode \breve{g}\else \u{g}\fi{}lu and A. Garg.
\newblock {\em Phys. Rev. Lett.}, 88:237205, 2002.

\bibitem{KececiogluGargB}
E. Ke\ifmmode \mbox{\c{c}}\else \c{c}\fi{}ecio\ifmmode \breve{g}\else \u{g}\fi{}lu and A. Garg.
\newblock {\em Phys. Rev. B}, 67:054406, 2003.

\bibitem{SeibergWitten}
N. Seiberg and E. Witten.
\newblock {\em Nucl. Phys. B}, 426:19, 1994.
\newblock Erratum-ibid. {\bf 430}, 485 (1994).

\bibitem{SeibergWittenB}
N. Seiberg and E. Witten.
\newblock {\em Nucl. Phys. B}, 431:484, 1994.

\bibitem{SWreview}
A. Bilal.
\newblock {\em "arxiv.org/abs/hep-th/9601007}, 1996.

\bibitem{TGMJPKAK}
T. Gulden, M. Janas, P. Koroteev and A. Kamenev.
\newblock {\em Sov. Phys. JETP}, 117:1, 2013.
\newblock [J. Exp. Theor. Fiz. 144:9, 2013].

\bibitem{TGMJAK}
T. Gulden, M. Janas and A. Kamenev.
\newblock {\em J. Phys. A: Math. Theor.}, 47:085001, 2014.

\bibitem{LiudelBarcoHill}
J. Liu, E. del Barco and S. Hill.
\newblock {\em A microscopic and spectroscopic view of quantum tunneling of magnetization}, in {\em Molecular Magnets: Physics and Applications, ed: J. Bartolome, F. Luis and J. F. Fernandez}, 12:77-110, 2014.

\bibitem{delBarco}
E. del Barco, A. D. Kent, S. Hill, J. M. North, N. S. Dalal, E. M. Rumberger, D. N. Hendrickson, N. Chakov and G. Christou.
\newblock {\em Jour. of Low Temp. Phys.}, 140:119-174, 2005.

\bibitem{Garanin}
D. A. Garanin and E. M. Chudnovsky.
\newblock {\em Phys. Rev. B}, 56:17, 1997.

\bibitem{Gatteschi}
D. Gatteschi, R. Sessoli and A. Cornia.
\newblock {\em Chem. Commun.}, 725-732, 2000.

\bibitem{WernsdorferSessoli}
W. Wernsdorfer and R. Sessoli.
\newblock {\em Science}, 284:133, 1999.

\bibitem{Barra}
A.-L. Barra, P. Debrunner, D. Gatteschi, C. E. Schulz and R. Sessoli.
\newblock {\em Europhys. Lett.}, 35:133-138, 1996.

\bibitem{Sangregorio}
C. Sangregorio, T. Ohm, C. Paulsen, R. Sessoli and D. Gatteschi.
\newblock {\em Phys. Rev. Lett.}, 78:24, 1997.

\bibitem{Garg}
A. Garg.
\newblock {\em Phys. Rev. B}, 60:9, 1999.

\bibitem{Bender2003}
C. M. Bender, D. C. Brody and H. F. Jones.
\newblock {\em Am. J. Phys.}, 71:1095-1102, 2003.

\bibitem{JoglekarSaxena}
Y. N. Joglekar and D. Scott and M. Babbey and A. Saxena.
\newblock {\em Phys. Rev. A}, 82:030103(R), 2010.

\bibitem{Nanayakkara}
A. Nanayakkara.
\newblock {\em J. Phys. A: Math. Gen.}, 37:4321-4334, 2004.

\bibitem{BenderClassTraj}
C. Bender and J.-H. Chen and D. Darg and K. Milton.
\newblock {\em J. Phys. A: Math. Gen.}, 39:4219-4238, 2006.

\end{thebibliography}
\end{document}